\documentclass[eqsecnum,twocolumn,aps,epsf]{revtex4}
\usepackage{graphicx}
\usepackage{color}
\begin{document}
\draft

\title{Hall coefficient in the ground state of stripe-ordered La$_{2-x}$Ba$_x$CuO$_4$ single crystals}

\author{Tadashi Adachi, Nobuyoshi Kitajima, and Yoji Koike}

\address{Department of Applied Physics, Graduate School of Engineering, Tohoku University, \\6-6-05 Aoba, Aramaki, Aoba-ku, Sendai 980-8579, Japan}

\date{\today}

\begin{abstract}
Temperature dependence of the Hall coefficient, $R_{\rm H}$, has been investigated in charge-spin stripe-ordered La-214 high-$T_{\rm c}$ superconductors. 
Using the simplest stripe-ordered system of La$_{2-x}$Ba$_x$CuO$_4$, it has been clarified that both the behavior of $R_{\rm H}$ and its sign exhibit significant dependences on the hole concentration. 
That is, $R_{\rm H}$ is zero in the ground state of the charge-spin stripe order at $x=1/8$, while it is negative in the less-stabilized state of the charge stripe for $x<1/8$. 
These are interpreted as owing to the delicate balance of the contributions of the hole-like Fermi surface and the possible electron pocket arising from the formation of the charge-spin stripe order. 

\end{abstract}
\vspace*{2em}
\pacs{PACS numbers: 74.25.F-, 74.62.Dh, 74.72.Gh}
\maketitle
\newpage

%*****************************************************************************************
%\section{Introduction}\label{intro}
%*****************************************************************************************
The Hall coefficient, $R_{\rm H}$, in the high-$T_{\rm c}$ superconducting (SC) cuprates has attracted considerable attention owing to its peculiar behavior since the early stages of high-$T_{\rm c}$ research. 
The $R_{\rm H}$ in the normal state is strongly dependent on temperature, which is unusual in conventional metallic superconductors. 
The temperature dependence of $R_{\rm H}$ at high temperatures has been explained based upon the Fermi-liquid model~\cite{kontani} or the two-carrier model~\cite{ono} taking into account spin fluctuations or charge fluctuations, respectively, but the behavior at low temperatures has not yet been understood. 

The experimental results of $R_{\rm H}$ at low temperatures in the La-214 cuprates such as La$_{2-x}$Ba$_x$CuO$_4$ (LBCO)~\cite{sera,ada-prb,ada-jpcs} and La$_{1.6-x}$Nd$_{0.4}$Sr$_x$CuO$_4$ (LNSCO)~\cite{noda} at $x \sim 1/8$ have revealed that $R_{\rm H}$ markedly decreases with decreasing temperature below the structural-phase-transition temperature between the tetragonal low-temperature (TLT) phase (space group: $P4_2/ncm$) and the orthorhombic mid-temperature (OMT) phase ({\it Bmab}), $T_{\rm d2}$. 
This has been explained as owing to the disappearance of the Hall voltage in the one-dimensional (1D) charge stripe-ordered state~\cite{tranquada} stabilized through the structural phase transition~\cite{noda} or owing to the cancellation of the Hall voltage by equal numbers of holes and electrons in the charge domain in the stripe-ordered state,~\cite{prelovsek} although it is still controversial. 
A gradual decrease in $R_{\rm H}$ in the normal state with decreasing temperature at low temperatures has been observed in YBa$_2$Cu$_3$O$_{7-\delta}$ (YBCO) with $\delta = 0.15 - 0.40$ (Ref. \cite{segawa}) and slightly Zn-substituted La$_{2-x}$Sr$_x$Cu$_{1-y}$Zn$_y$O$_4$ with $x=0.115$ and 0.15,~\cite{ada-jltp} which has been discussed in relation to the formation of the charge stripe order. 

Recently, measurements of $R_{\rm H}$ in high magnetic fields for underdoped YBCO with the hole concentration per Cu in the CuO$_2$ plane, $p=0.10-0.14$ by LeBoeuf {\it et al}.~\cite{leboeuf} have revealed a sign change of $R_{\rm H}$ at low temperatures.
They have proposed that the negative $R_{\rm H}$ originates from the formation of an electron pocket through the reconstruction of the Fermi surface caused by the possible formation of the charge stripe order. 
In YBCO, however, effects of the possible charge stripe order on various physical properties are relatively weak compared with those in the La-214 cuprates, preventing one from understanding explicitly the relation between the electron pocket and the charge stripe order. 

In this Rapid Communication, $R_{\rm H}$ in the charge stripe-ordered state is investigated in La-214 cuprates with various values of $p$. 
It has been found in LBCO with $x=0.08-0.12$ and LNSCO with $x=0.12$ that the behavior of $R_{\rm H}$ including the sign at low temperatures exhibits a significant dependence on $p$. 
That is, $R_{\rm H}$ for $x=0.10-0.12$ undergoing the phase transition to the TLT phase markedly decreases with decreasing temperature below $T_{\rm d2}$. 
Moreover, a sign change of $R_{\rm H}$ is observed at low temperatures for $x=0.10$ and the absolute value of the negative $R_{\rm H}$ decreases with increasing $x$, followed by almost zero for $x=0.12$ where the charge stripe order is completely stabilized. 
These results indicate that $R_{\rm H}$ is zero in the ground state of the charge-spin stripe order at $x \sim 1/8$ and that $R_{\rm H}$ becomes negative in the less-stabilized state of the charge stripe. 
It appears that there exists a close correlation between the stability of the charge stripe order, the sign of $R_{\rm H}$ and the topology of the Fermi surface discussed later.

%*****************************************************************************************
%\section{Experimental details}
%*****************************************************************************************
Single crystals of LBCO with $x=0.08$, 0.10, 0.11, 0.12 and LNSCO with $x=0.12$ were grown by the traveling-solvent floating-zone method. 
The detailed procedures are described elsewhere.~\cite{ada-prb,ada-prbmag} 
The composition of each crystal was analyzed by the inductively coupled-plasma (ICP) analysis. 
For LBCO and LNSCO with $x=0.10-0.12$, the charge stripe order is formed at low temperatures below $T_{\rm d2}$,~\cite{tranquada,fujita,fujita-physc} while it is not for LBCO with $x=0.08$.~\cite{fujita-physc} 
Both $R_{\rm H}$ and the ab-plane electrical resistivity, $\rho_{\rm ab}$, were measured by the standard ac six-probe method in magnetic fields parallel to the c-axis up to 9 T, by using a commercial apparatus (Quantum Design, PPMS). 
Both temperature and magnetic-field dependences of the Hall voltage were measured, resulting in good agreement of values of $R_{\rm H}$ with each other.

%*****************************************************************************************
%\section{Results}
%*****************************************************************************************
\begin{figure}[tbp]
\begin{center}
\includegraphics[width=1.0\linewidth]{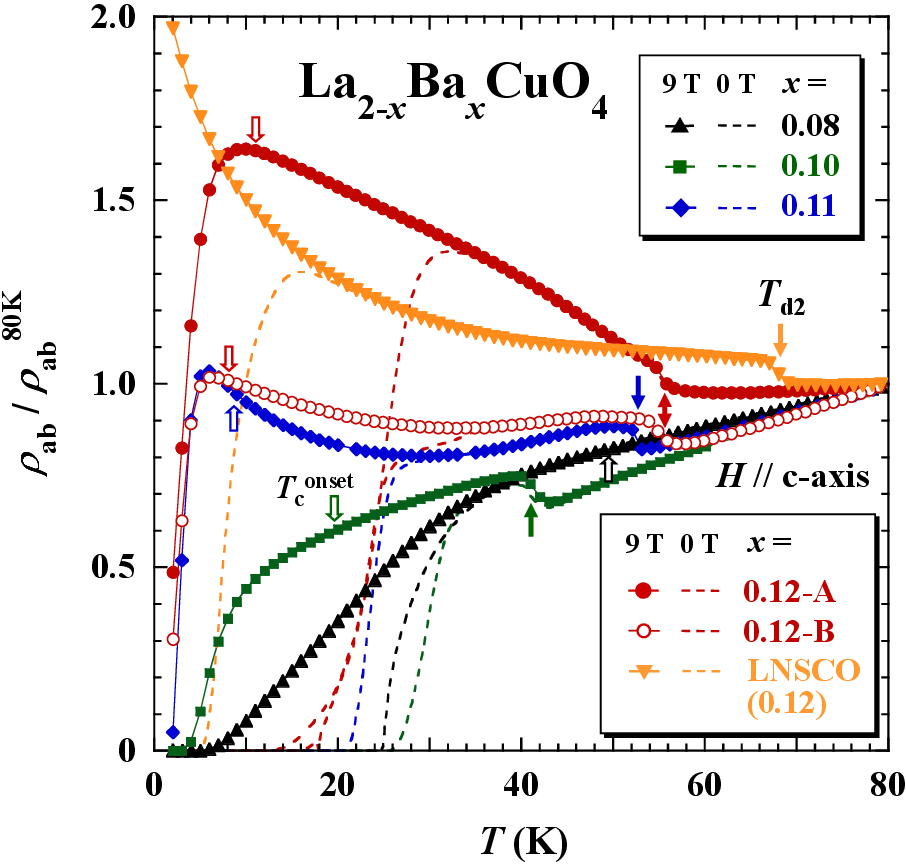}
\end{center}
\caption{(Color online) Temperature dependence of the ab-plane electrical resistivity, $\rho_{\rm ab}$, in zero field and a magnetic field of 9 T parallel to the c-axis normalized by its value at 80 K, $\rho_{\rm ab}^{\rm 80K}$, for La$_{2-x}$Ba$_x$CuO$_4$ with $x=0.08 - 0.12$ and La$_{1.6-x}$Nd$_{0.4}$Sr$_x$CuO$_4$ (LNSCO) with $x=0.12$. Note for La$_{2-x}$Ba$_x$CuO$_4$ with $x=0.12$ that the descriptions 'A' and 'B' correspond to different batches of the crystal. Solid arrows indicate the structural-phase-transition temperature between the TLT and OMT phases, $T_{\rm d2}$. Open arrows indicate the onset temperature of the SC transition, $T_{\rm c}^{\rm onset}$, in 9 T.} 
\label{p-d} 
\end{figure}

Figure 1 shows the temperature dependence of $\rho_{\rm ab}$ in zero field and a magnetic field of 9 T parallel to the c-axis normalized by its value at 80 K for LBCO and LNSCO with $x=0.08-0.12$. 
A jump in $\rho_{\rm ab}$ owing to the structural phase transition to the TLT phase is observed for $x=0.10-0.12$, as shown by solid arrows. 
For LBCO, $T_{\rm d2}$ systematically increases with an increase in $x$, the values of which are almost in agreement with those formerly reported.~\cite{suzuki} 
In zero field, $\rho_{\rm ab}$ exhibits a metallic behavior below $T_{\rm d2}$ for $x=0.10$ and 0.11, whereas it is less metallic or semiconducting for $x=0.12$. 
The behavior of $\rho_{\rm ab}$ for $x=0.12$ is typical of the sample with $p \sim 1/8$ where the charge stripe order is stabilized. 
For $x=0.12$ in LBCO, $\rho_{\rm ab}$ gradually decreases with decreasing temperature roughly below 30 K and goes to zero, which may be owing to a tiny inclusion of regions of $x<0.12$ having higher SC transition temperatures, $T_{\rm c}$'s, in the sample. 
It is also probable that the different behavior of $\rho_{\rm ab}$ between samples 'A' and 'B' of $x=0.12$ is owing to the different amount of a tiny inclusion of $x<0.12$. 
From the magnetic susceptibility measurements, in fact, $T_{\rm c}$, defined as the cross point between the extrapolated line of the steepest part of the shielding diamagnetism and zero susceptibility, is estimated to be as low as 4.2 K, indicating that most regions in the sample are made of $x=0.12$. 
It is noted that the present behavior of $\rho_{\rm ab}$ for $x=0.12$ in LBCO is different from that reported by Li {\it et al.},~\cite{li} because their reported $\rho_{\rm ab}$ in zero field is metallic even below $T_{\rm d2}$, as it suddenly drops at $\sim 40$ K and decreases gradually toward zero with decreasing temperature. 

\begin{figure}[tbp]
\begin{center}
\includegraphics[width=1.0\linewidth]{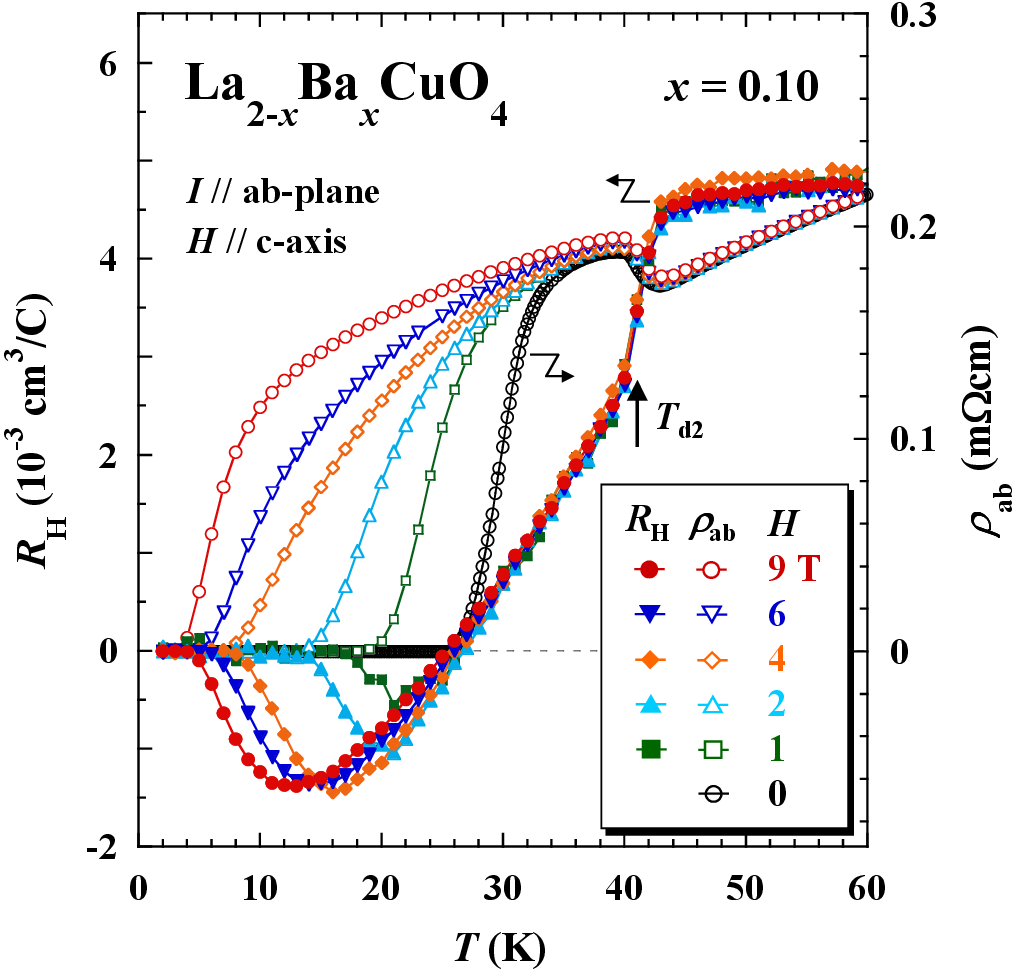}
\end{center}
\caption{(Color online) Temperature dependences of the Hall coefficient, $R_{\rm H}$, (left axis) and the ab-plane electrical resistivity, $\rho_{\rm ab}$, (right axis) in various magnetic fields parallel to the c-axis up to 9 T for La$_{2-x}$Ba$_x$CuO$_4$ with $x=0.10$. The arrow indicates the structural-phase-transition temperature between the TLT and OMT phases, $T_{\rm d2}$.} 
\label{p-d} 
\end{figure}

Temperature dependences of $R_{\rm H}$ and $\rho_{\rm ab}$ in various magnetic fields up to 9 T are displayed in Fig. 2 for LBCO with $x=0.10$. 
It is noted that similar temperature dependence of $R_{\rm H}$ is observed for LBCO with $x=0.11$ (Ref. \cite{ada-jpcs}).
With increasing field, the SC transition curve in $\rho_{\rm ab}$ vs. $T$ exhibits broadening characteristic of the underdoped high-$T_{\rm c}$ cuprates.~\cite{ada-prbmag} 
The $R_{\rm H}$ is almost independent of temperature above $T_{\rm d2}$, whereas it suddenly decreases below $T_{\rm d2}$. 
Moreover, a sign change of $R_{\rm H}$ is observed below $\sim 26$ K and eventually $R_{\rm H}$ goes to zero at low temperatures. 
Because temperatures at which $R_{\rm H}$ and $\rho_{\rm ab}$ become zero are in good agreement with each other, the behavior of $R_{\rm H}$ going to zero at low temperatures is owing to the SC transition. 
Accordingly, it is concluded that the sign of $R_{\rm H}$ in the ground state of LBCO with $x=0.10$ is negative. 

\begin{figure}[tbp]
\begin{center}
\includegraphics[width=1.0\linewidth]{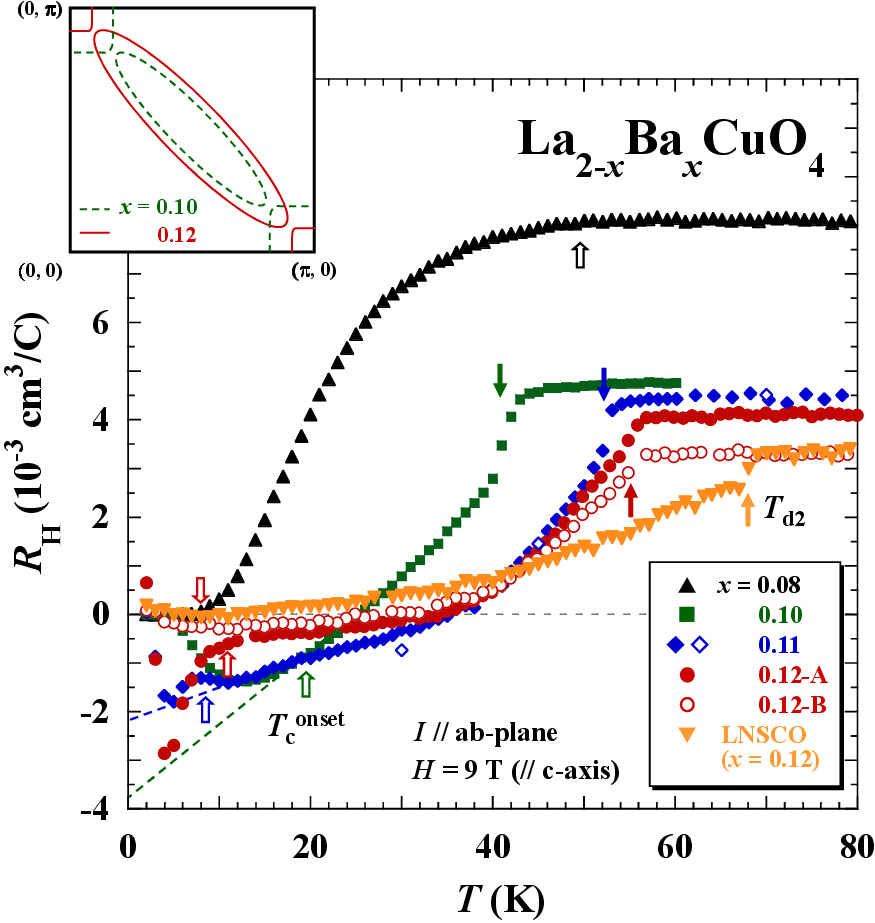}
\end{center}
\caption{(Color online) Temperature dependence of the Hall coefficient, $R_{\rm H}$, in 9 T parallel to the c-axis for La$_{2-x}$Ba$_x$CuO$_4$ (LBCO) with $x=0.08-0.12$ and La$_{1.6-x}$Nd$_{0.4}$Sr$_x$CuO$_4$ (LNSCO) with $x=0.12$. Note for LBCO with $x=0.12$ that the descriptions 'A' and 'B' correspond to different batches of the crystal. Solid arrows indicate the structural-phase-transition temperature between the TLT and OMT phases, $T_{\rm d2}$. Open arrows indicate the onset temperature of the SC transition, $T_{\rm c}^{\rm onset}$, in 9 T estimated from $\rho_{\rm ab}$ shown in Fig. 1. Open diamonds show data of $R_{\rm H}$ obtained from the measurements of the magnetic-field dependence of the Hall voltage for LBCO with $x=0.11$. The inset is a schematic drawing of one possible reconstruction of the Fermi surface through the formation of a commensurate AF order or a $d$DW order~\cite{chakravarty} for LBCO with $x=0.10$ and 0.12.}  
\label{p-d} 
\end{figure}

As for the $p$-dependent behavior of $R_{\rm H}$ in 9 T parallel to the c-axis shown in Fig. 3, $R_{\rm H}$ is found to be positive and it decreases with increasing $x$ at high temperatures $\sim 80$ K, as in the case of La$_{2-x}$Sr$_x$CuO$_4$ (LSCO) single crystals.~\cite{ono}
For $x=0.08$, $R_{\rm H}$ above the onset temperature of the SC transition, $T_{\rm c}^{\rm onset}$, estimated from $\rho_{\rm ab}$ in 9 T shown in Fig. 1, is almost independent of temperature, whereas it gradually decreases with decreasing temperature below $T_{\rm c}^{\rm onset}$ owing to the SC transition. 
For $x=0.11$ and 0.12, on the other hand, $R_{\rm H}$ markedly decreases below $T_{\rm d2}$ as well as for $x=0.10$ shown in Fig. 2. 
Because $T_{\rm c}^{\rm onset}$ is far below $T_{\rm d2}$, the marked decrease in $R_{\rm H}$ is irrespective of the SC transition. 
Comparing the data of $x=0.10$ with those of $x=0.11$, the extrapolation of the data above $T_{\rm c}^{\rm onset}$ to zero temperature suggests that the sign change of $R_{\rm H}$ is more marked for $x=0.10$ than for 0.11, as shown by the broken lines in Fig. 3. 
For LBCO and LNSCO with $x=0.12$, on the other hand, $R_{\rm H}$ becomes almost zero at low temperatures. 
These results are summarized as follows. 
For the samples in the TLT phase and therefore in the charge stripe-ordered state, $R_{\rm H}$ markedly decreases with decreasing temperature below $T_{\rm d2}$. 
Moreover, the sign change of $R_{\rm H}$ at low temperatures exhibits a significant dependence on $p$.

%*****************************************************************************************
%\section{Discussion}
%*****************************************************************************************
First, the marked decrease in $R_{\rm H}$ below $T_{\rm d2}$ is discussed. 
The sudden decrease in $R_{\rm H}$ accompanied by the formation of the charge stripe order has been observed formerly in LBCO~\cite{sera,ada-prb,ada-jpcs} and LNSCO.~\cite{noda} 
These former results of $R_{\rm H}$ are compatible with the present ones. 
A possible origin of the decrease in $R_{\rm H}$ together with the formation of the charge stripe order has been proposed by Noda {\it et al}.~\cite{noda} to be the disappearance of the Hall voltage owing to the formation of a 1D charge domain in the stripe-ordered state. 
From direct numerical calculations including a stripe potential within the $t$-$J$ model, on the other hand, Prelov\v{s}ek {\it et al.}~\cite{prelovsek} have suggested that equal numbers of holes and electrons owing to the half occupancy of holes in the charge domain leads to the disappearance of the Hall voltage. 
For $x=0.10$ and 0.11, where the half occupancy in the 1D charge domain is guessed to be maintained,~\cite{fujita-physc} however, $R_{\rm H}$ exhibits negative finite values at low temperatures, as shown in Fig. 3. 
Therefore, these two ways of thinking are compatible with the decrease $R_{\rm H}$ below $T_{\rm d2}$, while those are not enough to explain the negative $R_{\rm H}$ at low temperatures in the stripe-ordered state. 

The sign change of $R_{\rm H}$ is often observed in the SC fluctuation regime, which is understood to be owing to the vortex motion. 
In this case, it has been suggested that the application of magnetic field tends to suppress the sign change.~\cite{matsuda} 
As shown in Fig. 2, however, the sign change is not suppressed but enhanced by the application of magnetic field, and, moreover, $R_{\rm H}$ is negative even in the normal state above $T_{\rm c}^{\rm onset}$ for $x=0.10$ and 0.11.~\cite{sc-fluc} 
Therefore, the vortex motion is not the origin of the sign change of $R_{\rm H}$ in LBCO with $x=0.10$ and 0.11. 
Simply thinking, the sign of $R_{\rm H}$ is sensitive to the subtle curvature of the Fermi surface. 
As mentioned before, recently, LeBoeuf {\it et al.}~\cite{leboeuf} have found a pronounced sign change of $R_{\rm H}$ at low temperatures in strong magnetic fields in underdoped YBCO with $p=0.10-0.14$. 
Because quantum oscillations have been observed in YBCO with $p=0.10$,~\cite{leyraud} they have suggested that the negative $R_{\rm H}$ is a product of the formation of an electron pocket through the Fermi-surface reconstruction caused by the possible formation of the charge stripe order. 
A similar behavior of $R_{\rm H}$ has formerly been found in 2H-TaSe$_2$ where both a strong decrease and a sign change of $R_{\rm H}$ are observed accompanied by the transition to the charge-density-wave (CDW) state.~\cite{lee} 
These have been explained theoretically in terms of the Fermi-surface reconstruction owing to the opening of the CDW gap.~\cite{evtushinsky} 
Accordingly, the sign change of $R_{\rm H}$ at low temperatures in LBCO with $x=0.10$ and 0.11 is possibly explained as owing to the creation of an electron pocket on the Fermi surface caused by the formation of the charge stripe order. 

Then, why does the behavior of $R_{\rm H}$ at low temperatures depend significantly on $p$? 
Supposed that the Fermi surface in the underdoped cuprates is reconstructed through the formation of a commensurate antiferromagnetic (AF) order or a $d$-density-wave ($d$DW) order,~\cite{chakravarty} both hole pockets and an electron pocket are created on the Fermi surface, located around ($\pm \pi/2$, $\pm \pi/2$) and ($\pm \pi$, 0), (0, $\pm \pi$) in the reciprocal lattice space, respectively, which is schematically shown in the inset of Fig. 3. 
Actually, recent angle-resolved photoemission experiments have revealed a pocket around ($\pi/2$, $\pi/2$).~\cite{chang,chinese} 
Considering the $p$-dependent change of the Fermi-surface topology,~\cite{ino} the possible electron pocket around ($\pm \pi$, 0), (0, $\pm \pi$) tends to shrink with increasing $p$, as shown in the inset of Fig. 3. 
This is naturally suggestive of the weakening of the electron aspect and the development of the hole aspect on $R_{\rm H}$ with increasing $p$. 
This picture is consistent with the present results that the sign change of $R_{\rm H}$ gradually weakens and $R_{\rm H}$ at low temperatures becomes zero with increasing $x$ from $x=0.10$ to 0.12. 
It is noted that $R_{\rm H}$ is positive at low temperatures for LNSCO with $x=0.15$,~\cite{noda} which is also consistent with this picture. 
Accordingly, the $p$-dependent change of $R_{\rm H}$ in the ground state is able to be understood by the delicate balance of the contributions of hole and electron pockets.

Millis {\it et al.}~\cite{millis} and Lin {\it et al.}~\cite{lin} have investigated theoretically the Fermi-surface reconstruction within the tight-binding model, including the charge and spin stripe potential. 
According to their calculations, a rather complicated arrangement of electron pockets and hole pockets is reconstructed and the development of the spin stripe correlation makes $R_{\rm H}$ negative, while the development of the charge stripe correlation makes $R_{\rm H}$ positive. 
That is, in the charge-spin stripe-ordered state, the sign of $R_{\rm H}$ depends on the delicate balance of the developments of the charge and spin stripe order. 
Based upon their calculations, it follows that $R_{\rm H}$ becomes zero owing to complete developments of both charge and spin stripe order at $x=1/8$, and that $R_{\rm H}$ becomes negative owing to the incomplete development of the charge stripe order for $x<1/8$.~\cite{fujita-physc} 
Actually, the sign change of $R_{\rm H}$ has been observed in YBCO with $p=0.10-0.14$ (Ref.~\cite{leboeuf}) and LSCO with $x=0.12$ (Ref.~\cite{suzuki-lysco}) where the spin stripe correlation is developed but the charge stripe order is not. 
Accordingly, it may be the case that $R_{\rm H}$ becomes zero owing to the stabilization of the charge-spin stripe order at $x=1/8$, while it becomes negative owing to the instability of the charge stripe order for $x<1/8$. 

Finally, we comment on the value of $R_{\rm H}$ in the ground state of LBCO. 
For $x=0.10$, the extrapolated value of $R_{\rm H}$ to zero temperature is estimated from Fig. 3 to be $-0.00375$ cm$^3$/C. 
Thus, based on the simple one-carrier model, the carrier density $n_{\rm Hall}$ is calculated to be $n_{\rm Hall} = - V_{\rm cell} / eR_{\rm H} / 2 = 0.0795$ per Cu in the CuO$_2$ plane, where $V_{\rm cell}$ is the volume of the unit cell in the notation of the tetragonal at high temperature ($I4/mmm$). 
For YBCO with $p=0.10$,~\cite{leboeuf} on the other hand, it has been reported that $n_{\rm Hall} = 0.0145$ per Cu in the CuO$_2$ plane. 
This value is comparable to the deduced value of 0.019 per Cu in the CuO$_2$ plane from the quantum-oscillation experiments.~\cite{leyraud} 
Assuming the negative value of $R_{\rm H}$ originates from a possible pocket on the Fermi surface, a simple comparison of $n_{\rm Hall}$ between LBCO and YBCO results in a larger pocket in LBCO than in YBCO. 
This appears to be inconsistent with that expected from the experimentally observed Fermi-surface topology,~\cite{ino,nakayama} because the reconstructed Fermi surface owing to the commensurate AF order or the $d$DW order is expected to produce a larger electron pocket around $(\pm \pi,0)$, $(0,\pm \pi)$ in YBCO than in LBCO. 
Accordingly, it is hard to have a quantitative discussion on the negative value of $R_{\rm H}$ in the ground state by using the simple one-carrier model.

%*****************************************************************************************
%\section{Summary}
%*****************************************************************************************
In conclusion, $R_{\rm H}$ in the stripe-ordered LBCO and LNSCO in the TLT phase markedly decreases owing to the formation of the charge-spin stripe order. 
In the ground state, on the other hand, $R_{\rm H}$ is zero in the completely ordered charge-spin stripe state at $x=1/8$, while it is negative in the less-stabilized state of the charge stripe for $x<1/8$. 
The $p$-dependent behavior of $R_{\rm H}$ including its sign is interpreted as owing to the delicate balance of the contributions of the holelike Fermi surface and the possible electron pocket arising from the formation of the charge-spin stripe order.

%*****************************************************************************************
%\section*{Acknowledgments}
%*****************************************************************************************
Fruitful discussions with T. Tohyama are gratefully acknowledged. 
We are indebted to M. Ishikuro for his help in the ICP analysis. 
This work was supported by a Grant-in-Aid for Scientific Research from the Ministry of Education, Science, Sports, Culture and Technology, Japan.


\begin{references}
\bibitem{sc-fluc}
It is found in Fig. 1 that the SC fluctuation in 9 T is suppressed down to $\sim 20$ K for $x=0.10$ and $\sim 8$ K for $x=0.11$ owing to the formation of the stripe order in the magnetic field,~\cite{ada-prbmag} suggesting the occurrence of the sign change of $R_{\rm H}$ at $\sim 26$ K for $x=0.10$ and 0.11~\cite{ada-jpcs} in the true normal state.
\end{references}
\end{document}